\documentclass[12pt]{article}
\usepackage{amsmath}
\usepackage{graphicx,psfrag,epsf}
\usepackage{enumerate}
\usepackage{natbib}
\usepackage{hyperref}
\newcommand{\blind}{0}
\def\code#1{\texttt{#1}}
\usepackage{graphics}

\usepackage{adjustbox}
\usepackage{amssymb}
\usepackage{bm}
\usepackage{graphicx}
\usepackage{natbib}
\usepackage{amsmath}
\makeatletter
\newcommand{\distas}[1]{\mathbin{\overset{#1}{\kern\z@\sim}}}%
\newsavebox{\mybox}\newsavebox{\mysim}
\newcommand{\distras}[1]{%
  \savebox{\mybox}{\hbox{\kern3pt$\scriptstyle#1$\kern3pt}}%
  \savebox{\mysim}{\hbox{$\sim$}}%
  \mathbin{\overset{#1}{\kern\z@\resizebox{\wd\mybox}{\ht\mysim}{$\sim$}}}%
}
\usepackage{breqn}
\usepackage{multirow}
\usepackage{bm}
\usepackage{multirow}
\usepackage[normalem]{ulem}
\useunder{\uline}{\ul}{}
\usepackage{graphicx}
\usepackage{wrapfig}
\usepackage{lscape}
\usepackage{rotating}
\usepackage{epstopdf}
\usepackage{comment}

\usepackage[utf8]{inputenc}
\usepackage[english]{babel}
 \usepackage{amsthm}

\usepackage{graphicx}
\usepackage{subcaption}

\usepackage{booktabs,subcaption,amsfonts,dcolumn}

\newcommand{\iid}{\stackrel{iid}{\sim}}

\usepackage{comment}

\begin{document}

\def\spacingset#1{\renewcommand{\baselinestretch}%
{#1}\small\normalsize} \spacingset{1}


\if0\blind
{
  \title{\bf Probabilistic Diffusion MRI Fiber Tracking Using a Directed Acyclic Graph Auto-Regressive Model of Positive Definite Matrices}

  \author{
  Zhou Lan\\
    Department of Statistics, North Carolina State University\\
    and \\
    Brian J Reich \\
    Department of Statistics, North Carolina State University\\
    }

  \maketitle
} \fi

\if1\blind
{
  \bigskip
  \bigskip
  \bigskip
  \begin{center}
    {\LARGE\bf Title}
\end{center}
  \medskip
} \fi

\bigskip
\begin{abstract}
Diffusion MRI is a neuroimaging technique measuring the anatomical structure of tissues. Using diffusion MRI to construct the connections of tissues, known as fiber tracking, is one of the most important uses of diffusion MRI. Many techniques are available recently but few properly quantify statistical uncertainties. In this paper, we propose a directed acyclic graph auto-regressive model of positive definite matrices and apply a probabilistic fiber tracking algorithm. We use both real data analysis and numerical studies to demonstrate our proposal.
\end{abstract}

\noindent%

\spacingset{1.5}

\section{Introduction}
Among the many uses of diffusion MRI, using tractography methods to depict the underlying white matter fiber tracts of tissues may be the most important. Procedures of identifying tracts are referred as to \textit{fiber tracking}. For clinical practice, fiber tracking provides potential benefits for presurgical planning \citep{chung2011principles}. In neuroscience, understanding the anatomical connection of the brain is an important component of the connectome \citep{sporns2005human}.

Many technologies have been developed for fiber tracking in recent years. Among these methods, \code{DiST} (short for \textit{Diffusion Direction Smoothing and Tracking}) \citep{wong2016fiber} is one of the most prominent \citep{kang2016discussion,schwartzman2016discussion,lazar2016discussion}. \code{DiST} is composed of three major steps: In Step 1, voxel-wise diffusion directions are estimated using diffusion-weighted signals; In Step 2, the estimated diffusion directions obtained in Step 1 are smoothed over space; Finally in Step 3, the smoothed diffusion directions are taken as the inputs of a fiber tracking algorithm that determines if some voxels construct a fiber.

Although \code{DiST} has many appealing features, it has some limitations \citep{kang2016discussion,lazar2016discussion,schwartzman2016discussion}. A major limitation may be that the separate steps of \code{DiST} make it difficult to properly account for statistical uncertainties. In light of this, we propose a Bayesian hierarchical approach which allows valid statistical inference for fiber tracking. First, we assume that the logarithm signals follow a normal distribution, simplifying the model by avoiding the challenging Rician distribution \citep{wong2016fiber}. Also, we induce spatial smoothness using a random field for spatially dependent positive definite matrices instead of the optimization-based smoothing procedure of \code{DiST}. This avoids the optimization issue raised by Schwartzman \citep{schwartzman2016discussion}. 

In the rest of the paper, we first introduce our method, referred as to \code{SpDiST} (short for  \textit{Spatial Diffusion Direction Smoothing and Tracking}). To demonstrate our proposal, we use both real data analysis and a simulation study. The real data analysis demonstrates that our proposal provides a valid and efficient means to quantify the uncertainties of fiber tracking. Moreover, the simulation study shows that our proposal produces accurate estimation. Finally, we conclude with a discussion.

\section{Method: \code{SpDiST}}
\label{sec:method}

\subsection{Spatial Tensor Model}
\label{sec:step1}
In this section, we introduce the spatial tensor model based on directed acyclic graph auto-regression for positive definite matrices. The diffusion MRI has $m\in\{1,2,...,M\}$ measurements at voxel $v\in\{1,2,...,n\}$, denoted as $S_{mv}\in\mathbb{R}^+$. The measurements $S_{mv}$ are used to estimate the diffusion tensor $\bm{A}_v$ for voxel $v$. $\bm{A}_v$ is a $3\times 3$ positive definite matrix interpreted as covariance matrix of a local Brownian motion, indicating the local tensor direction. The goal is to use the measurements $S_{mv}$ to obtain tensor direction information from $\bm{A}_v$.

The noiseless signal intensity $\bar{S}_{mv}$ can be expressed in terms of $\bm{A}_v$  \citep{mori2007introduction} as $$\bar{S}_{mv}=S_{0v}\exp(-b\bm{g}_m^T\bm{A}_v\bm{g}_m).$$ In this expression, $S_{0v}$, $b$, and $\bm{g}_m$ are non-diffusion weighted intensity, scale parameter, and $3\times1$ unit-norm
gradient vector, respectively. A detailed explanation of these three quantities can be found in Soares et al. \citep{soares2013hitchhiker}. Given $\bm{A}_v$, $\bar{S}_{mv}$ can be understood as the probability intensity of the Gaussian motion when measuring at direction $\bm{g}_v$. For statistical modeling, $S_{0v}$, $b$, and $\bm{g}_m$ can simply be understood as fixed and known values.

The observations $S_{mv}$ are noisy realizations of $\bar{S}_{mv}$. The Rician distribution \citep{wong2016fiber} is reasonable for modeling $S_{mv}$ but it causes computational issues. Here, we assume that the noise is a multiplier to the $\bar{S}_{mv}$ and it follows a lognormal distribution. The model is
\begin{equation}
\label{eq:gaussian}
\begin{aligned}
    \log S_{mv}= \log S_{0v}-b\bm{g}_m^T\bm{A}_v\bm{g}_m+\epsilon_{mv},\quad\epsilon_{mv}\iid\mathcal{N}(0,\sigma^2),
    \end{aligned}
\end{equation}
where $\epsilon_{mv}$ is the noise following a mean-zero normal distribution with variance $\sigma^2$.

To induce spatial smoothness, an image is treated as a directed graph whose nodes are voxels and whose directed edges are from node $v$ to nodes in $N(v)$. Following Datta et al. \citep{datta2017spatial}, we use the directed acyclic graph (directed and no loops) to construct to $N(v)$, leading to a valid joint density function of $[\bm{A}_1, \bm{A}_2,..., \bm{A}_n]$. In particular, we assume that the conditional mean of $\bm{A}_v$ is the average of its neighboring tensors, denoted as $\mathbb{E}[\bm{A}_v|\bm{A}_u,u\in N(v)]=\frac{1}{|N(v)|}\sum_{u\in N(v)}\bm{A}_u$, where $N(v)$ is a set containing neighboring voxel indices of voxel $v$, and $|N(v)|$ is the set size.

In a directed acyclic graph, we have at least one voxel $v$ whose $N(v)$ is an empty set. For $N(v)$ is an empty set, we assume that $\bm{A}_{v}$ follows a Wishart distribution with mean matrix $\bm{I}$ and degrees of freedom $k$. Otherwise, conditional on $\bm{A}_u,u\in N(v)$, we assume that $\bm{A}_{v}$ follows a Wishart distribution with mean matrix $\bar{\bm{A}}_{v}=\frac{1}{|N(v)|}\sum_{u\in N(v)}\bm{A}_u$ and degrees of freedom $k$. The model is

\begin{equation}
\label{eq:rdf}
\begin{aligned}
    \bm{A}_{v}|\bm{A}_u,u\in N(v)&\sim\mathcal{W}\left(\bar{\bm{A}}_{v},k\right)&\quad& \text{if $N(v)$ is not empty},\\
    \bm{A}_{v}&\sim \mathcal{W}(\bm{I},k)&\quad& \text{if $N(v)$ is empty}.
\end{aligned}
\end{equation}
In Equation (2), to preserve the designed mean realizations, we parameterize the Wishart distribution for $\bm{X}\sim\mathcal{W}(\bm{V},k)$ to have $\mathbb{E}\bm{X}=\bm{V}$. The probability density function is $${\displaystyle f(\mathbf {X} )={\frac {|\mathbf {X} |^{(k-p-1)/2}e^{-\operatorname {tr} ([\mathbf {V}/k] ^{-1}\mathbf {X} )/2}}{2^{\frac {kp}{2}}|{\mathbf {V}/k }|^{k/2}\Gamma _{p}({\frac {k}{2}})}}},$$ where $p$ is the matrix dimension and $\Gamma _{p}({\frac {k}{2}})$ is the multivariate gamma function. 

\begin{figure}[t!]
    \centering
    \includegraphics[width=0.8\textwidth]{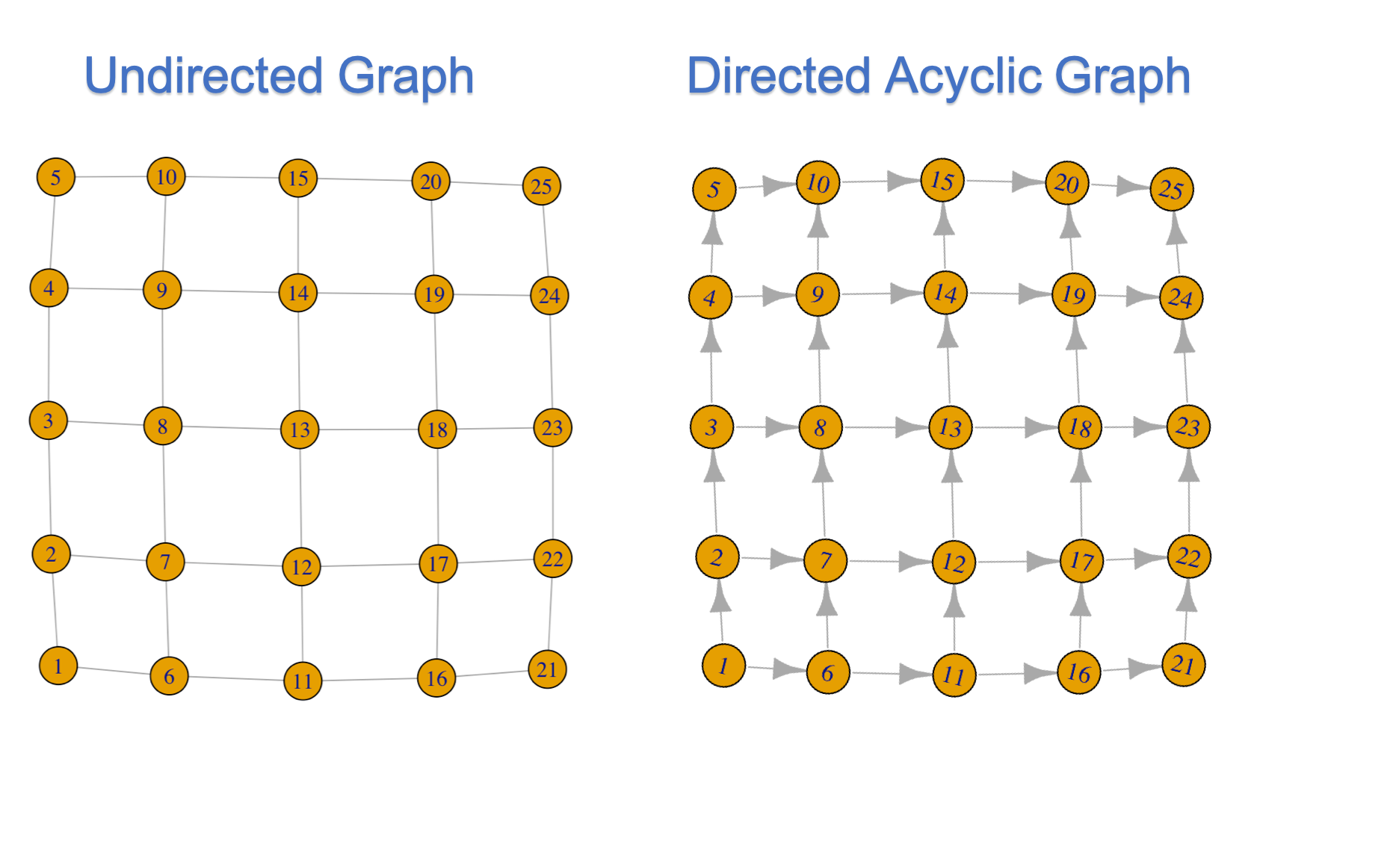}
    \caption{The construction of a directed acyclic graph based on an undirected graph. The left panel is the undirected graph of a $5\times5$ image. The right pannel is the corresponding directed acyclic graph after modifying the edges.}
    \label{fig:graph}
\end{figure}

Here, we give an approach to construct a directed acyclic graph. For an image, we construct an undirected graph whose voxels are nodes, and the neighboring nodes are connected. We order the voxels by their coordinates, i.e., for a 2D image on a \textit{x-y} axis, we first order the voxels according to their coordinates of the \textit{y}-axis, then next we order the voxels according to their coordinates of the \textit{x}-axis. For each edge of the undirected graph, we modify the undirected edge to a directed edge which is from the node with a smaller rank to a node with a larger rank. The modified graph is a directed acyclic graph whose edges connect neighboring voxels. In Figure \ref{fig:graph}, we give an example describing how a directed acyclic graph for a $5\times 5$ image is constructed.

\subsubsection{MCMC Algorithm}
We use MCMC for model fitting. We give $k\sim\mathcal{U}(3,50)$ and $\sigma^{-2}\sim\mathcal{GA}(0.01,0.01)$. The primary challenge in the MCMC algorithm is to sample the posterior of $\bm{A}_v$. Since the prior of $\bm{A}_v$ is not conjugate, we sample it using single-site Metropolis-Hastings sampling with Wishart distribution $\mathcal{W}(\bm{A}_v'|\bm{A}_v,q)$ as the proposal distribution. The algorithm is described below:

\begin{description}
    \item[Candidate Generation:] Generate a candidate sample $\bm{A}_v'$ using $\bm{A}_v'\sim\mathcal{W}(\bm{A}_v'|\bm{A}_v,q)$;
    \item[Acceptance Rate:] Calculate the acceptance rate $r(\bm{A}_v',\bm{A}_v)=\frac{\mathcal{L}(\bm{A}_v'|.)\mathcal{W}(\bm{A}_v|\bm{A}_v',q)}{\mathcal{L}(\bm{A}_v|.)\mathcal{W}(\bm{A}_v'|\bm{A}_v,q)}$, where \begin{equation}
    \begin{aligned}
        \mathcal{L}(\bm{A}_v^*|.)&\propto\prod_{m=1}^M\mathcal{N}(\log S_{mv}|\log S_{0v}-b\bm{g}_m^T\bm{A}_v^*\bm{g}_m,\sigma^2)\times\\
        &\mathcal{W}(\bm{A}_v^*|\bar{\bm{A}}_v^*,q)\prod_{u:v\in N(u)}\mathcal{W}(\bm{A}_u|\bar{\bm{A}}_u^*,q),\\
        \end{aligned}
    \end{equation}
    where $\bar{\bm{A}}_u^*=\frac{\sum_{u\in N(u)/v} \bm{A}_u+ \bm{A}_v^*}{|N(v)|}$. $\mathcal{W}(.|\bm{A},\nu)$ and $\mathcal{N}(.|\mu,\sigma^2)$ are the density functions of Wishart distribution and normal distribution, respectively.
    \item[Decision:] Generate $u\sim\mathcal{U}(0,1)$. If $u<r(\bm{A}_v',\bm{A}_v)$, accept $\bm{A}_v'$.
\end{description}
The acceptance rate can be tuned by the degrees of freedom $q$, where smaller $q$ leads to smaller acceptance rate. We tune $q$ to make the acceptance rate around $0.4$. 

We use Metropolis-Hastings algorithm with log-normal random walk as proposal distribution to update the degrees of freedom $q$ and use Gibbs sampling to update $\sigma^2$ based on its posterior $[\sigma^{-2}|.]\sim \mathcal{GA}(Mn/2+0.01,\sum_{m,v}(\log S_{mv}-\log S_{0v}+b\bm{g}_m^T\bm{A}_v\bm{g}_m)^2/2+0.01)$.

\subsection{Probabilistic Fiber Tracking Algorithm}
\label{sec:step2}
\begin{figure}[t!]
    \centering
    \includegraphics[width=0.5\textwidth]{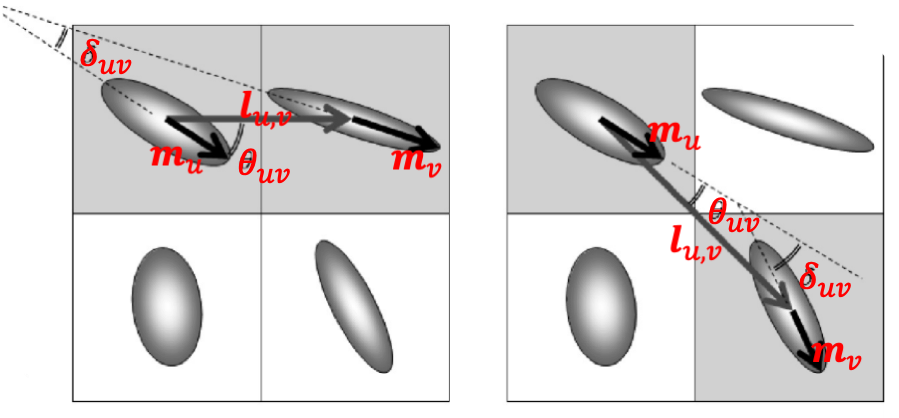}
    \caption{Two angles between two voxels. This figure is modified from Chung et al. \citep[][Figure 3]{chung2011principles}.}
    \label{fig:angle}
\end{figure}

We collect the $T$ MCMC samples of $\bm{A}_v$, denoted as $\{\bm{A}_v^{(t)}:t=1,2,...,T\}$. For each sample, we compute the principal eigenvector of $\bm{A}_v^{(t)}$, denoted as $\bm{m}_v^{(t)}$. For each posterior draw, we use $\bm{m}_v^{(t)}$ as inputs of a fiber tracking algorithm. In this paper, we continue to use the Fiber Assignment by Continuous Tracking (FACT) \citep{mori1999three}, following Wong et al. \citep{wong2016fiber}. The algorithm can be stated as 
\begin{itemize}
                \item \textbf{Initialization:} Starting from \textit{seed voxels};
                \item \textbf{Recursive:} Starting with voxel $u$, we search neighboring voxels and compute the two angles: $\delta_{uv}=\text{arccos}\left(\frac{\bm{m}_v^T\bm{m}_u}{|\bm{m}_v||\bm{m}_u|}\right)$ is the angle between the two tensor directions ($\bm{m}_u$ and $\bm{m}_u$) and $\theta_{uv}=\text{arccos}\left(\frac{\bm{m}_u^T\bm{l}_{u,v}}{|\bm{m}_u||\bm{l}_{u,v}|}\right)$ is the angle between the current tensor ($\bm{m}_u$) and between-voxel direction ($\bm{l}_{u,v}$). See Figure \ref{fig:angle} for details. We move to the voxels with $\theta<C$ and $\delta<C$. If there are multiple voxels statisfying this condition, we move to all the voxels and treat each voxel as a current voxel for next iteration.;
                \item \textbf{Result:} Sequences of voxels constructing \textit{fibers}.
            \end{itemize}
Since we apply the algorithm for each posterior draw, the algorithm returns $T$ possible fibers. We summarize $K$ distinct patterns from the outputs and calculate the associated probability for pattern $k\in\{1,2,...,K\}$ defined as $\frac{T_k}{T}$, where $T_k$ is the frequency of the pattern $k$. This procedure is known as \textit{probabilistic fiber tracking} and quantifies the uncertainties of fiber tracking result.

\section{Real Data Application}
\label{sec:real}
In this section, we use a real data example \citep[][Section 6]{dryden2009non} to demonstrate our proposal. In particular, we focus on uncertainty quantification. The real data has $50\times20$ voxels and $M=15$ measurements. A detailed description can be found in Dryden et al. \citep[][Section 6]{dryden2009non}. We sample $2000$ MCMC samples after discarding $3000$ samples as burn-in and thin the MCMC chain by retaining every $100$ iterations of the chain.

\begin{figure}
    \centering
    \includegraphics[width=1\textwidth]{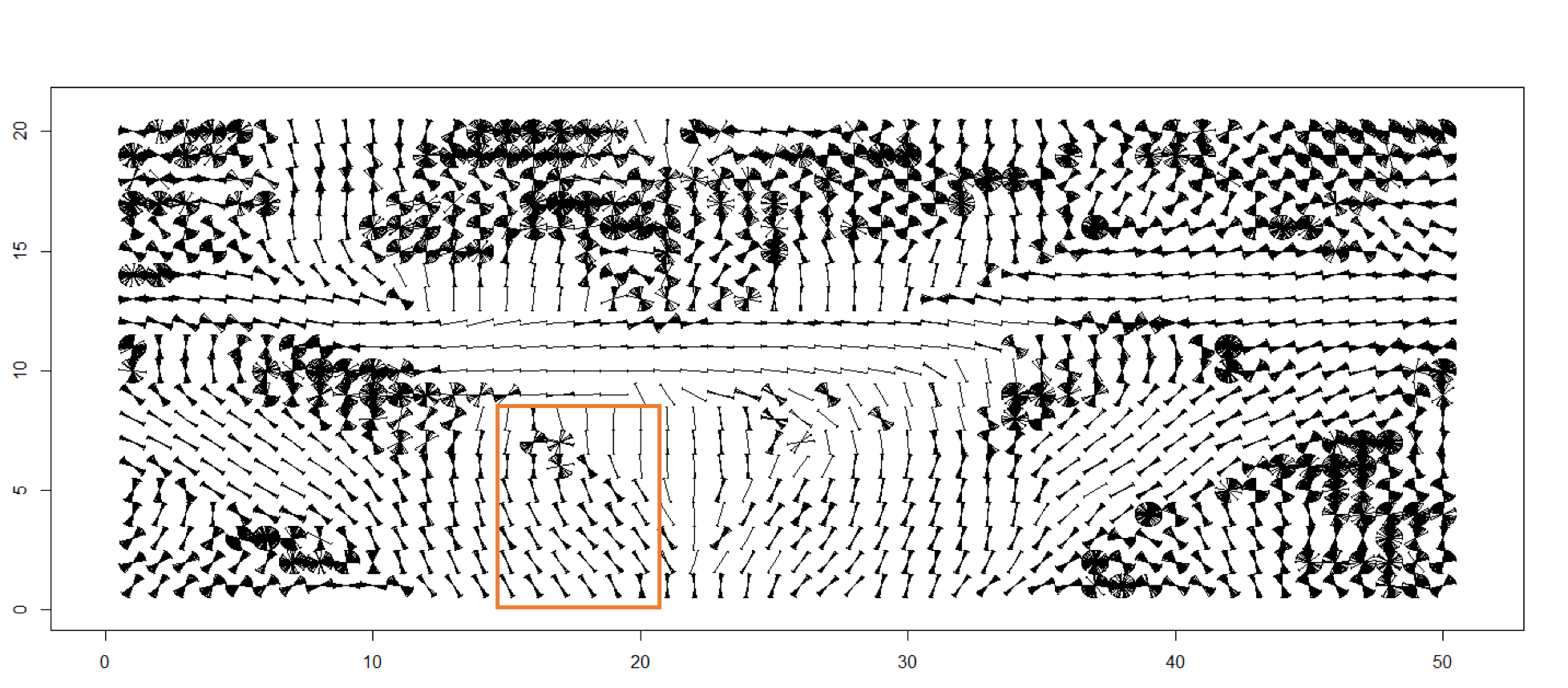}
    \caption{For each voxel, the MCMC samples of $\bm{m}_v$ are overlaid on the its location on a $50\times20$ map. For each voxel, we plot . The voxels with heterogeneous directions have large uncertainties. Otherwise, there are small uncertainties.}
    \label{fig:realfull}
\end{figure}

Since it is more efficient to visualize tensor directions in a 2D environment and the image is 2D, we focus on the first two dimensions of $\bm{A}_v$ and compute the corresponding principal eigenvector $\bm{m}_v$. To quantify the uncertainties of tensor direction estimation in each voxel, we overlay the MCMC samples on a $50\times20$ map (Figure \ref{fig:realfull}). In Figure \ref{fig:realfull}, the voxels with heterogeneous directions have large uncertainties. Otherwise, there are small uncertainties.

\begin{figure}[t!]
    \centering
    \includegraphics[width=1\textwidth]{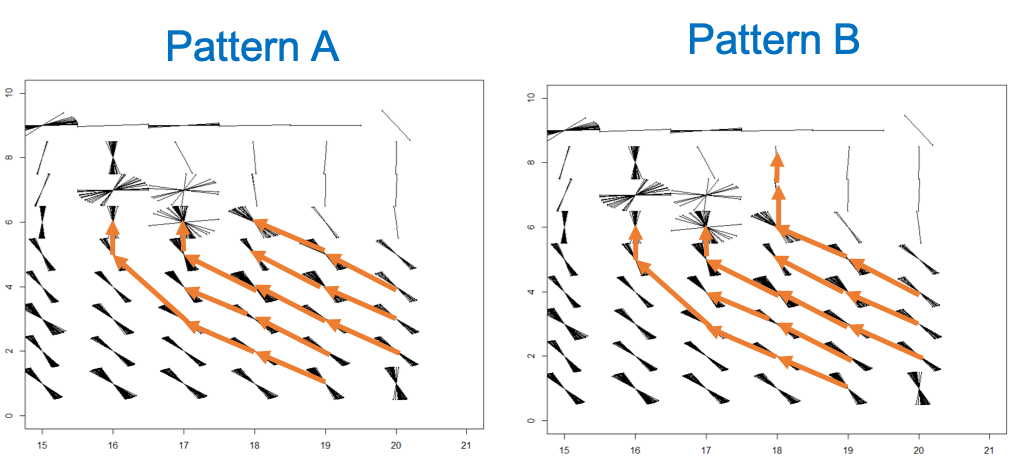}
    \caption{The consecutive orange arrows construct a fiber. Two patterns are identified, where the left pattern and the right pattern are denoted as Pattern A and Pattern B, respectively.}
    \label{fig:realpartial}
\end{figure}

Figure \ref{fig:realfull} only provides voxel-wise uncertainties but our fully-Bayesian approach can propagate spatial uncertainties through to uncertainty in fiber tracking. In this way, the MCMC-based \code{SpDiST} also provides a probabilistic approach to quantifying the uncertainties of fiber tracking. To have a concise and representative illustration, we focus on the region in the orange box of Figure \ref{fig:realfull}. We apply the FACT algorithm as described in Section \ref{sec:step2}. In light of the conventions in setting the threshold $C$ \citep{chung2011principles}, we consider $C$ ranging from $18^o$ to $28^o$. There are two distinct patterns (Pattern A and Pattern B) for $C\in [18^o, 28^o]$ (Figure  \ref{fig:realpartial}) as dominating the posterior probability of the tract. These two tracts differ only by how far the tract continues vertically in column 18. The probabilities of each pattern vary by different thresholds $C$. 

\begin{figure}[t!]
    \centering
    \includegraphics[width=1\textwidth]{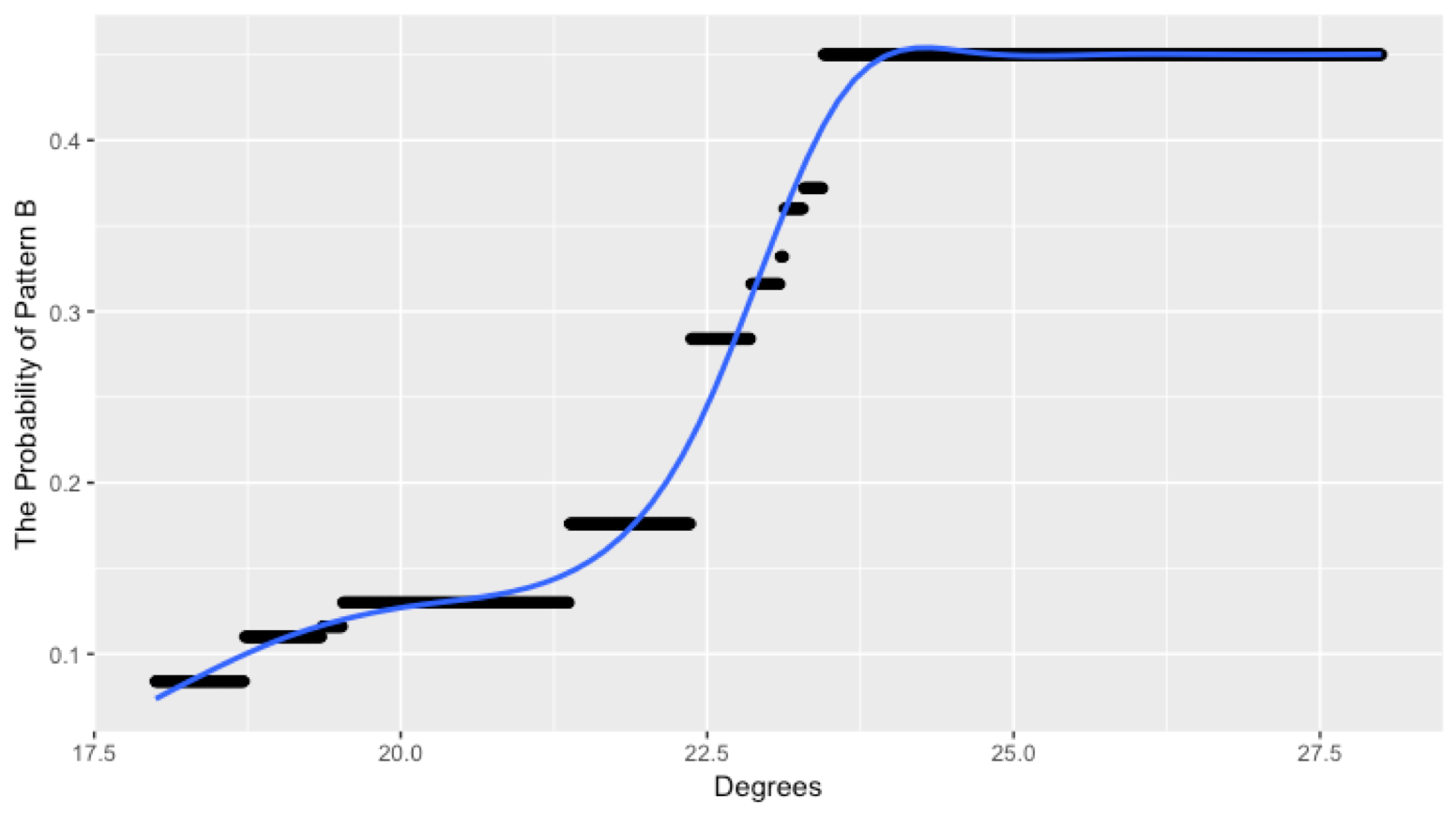}
    \caption{The probability of Pattern B varies with different threshold $C$. When $C$ is ranging from $24^o$ to $28^o$ the probability is insensitive to $C$.}
    \label{fig:sensitivity}
\end{figure}

Kang and Li \citep[][Section 3]{kang2016discussion} show that the FACT algorithm hinges on the tuning parameter $C$. It requires a sensitivity analysis to explore the impact of $C$. Here, we give a sensitivity analysis. We apply the FACT algorithm with $C=18+0.01\times s$ and $s=\{0,1,2,...,1000\}$. Since there are only two distinct patterns, we report the probabilities of Pattern B with different thresholds $C$ (Figure \ref{fig:sensitivity}). We find that the result is sensitive to the choice of $C$ unless it is ranging from $24^o$ to $28^o$.

\section{Numerical Study}
\label{sec:sim}
\subsection{Data Description}
\begin{figure}[t!]
    \centering
    \includegraphics[width=0.8\textwidth]{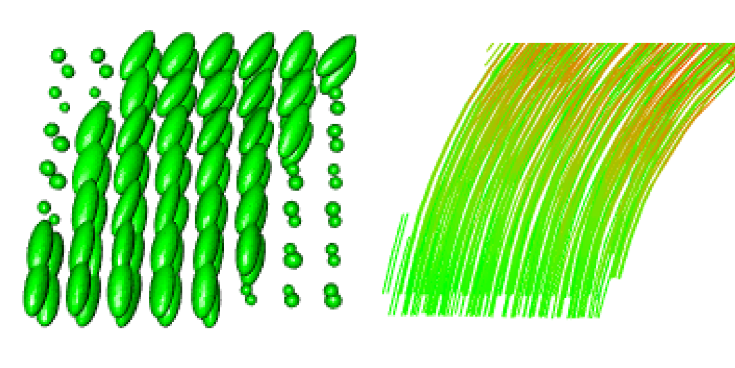}
    \caption{The tensor directions (left panel) and underlying fibers (right panel) of the example data .}
    \label{fig:sim}
\end{figure}

In this section, we use synthetic diffusion-weighted signals in Wong et al. \citep[][S6]{wong2016fiber} and further modify them for our numerical study. In total, we have $8\times7\times2$ voxels where the three digits represent the dimension of $x$-axis, $y$-axis, and $z$-axis, respectively. The underlying tensors and fibers from the synthetic signals are displayed in Figure \ref{fig:sim}. A comprehensive description of example data generation can be found in Wong et al. \citep[][S6]{wong2016fiber} (i.e., generating model, parameters, true tensor directions, etc.). Here, we give a brief description. The fibers are essentially arcs with the center point at right/left bottom points. For voxels composing fibers, its principal eigenvector $\bm{m}_v$ is tangent to the arc. The noiseless signal in the example data is given as $\bar{S}_{mv}=S_{0v}\exp[-b(\bm{g}_m^T\bm{m}_v)^2]$, a reparameterized model of Model \ref{eq:gaussian} \citep{wong2016fiber}.

To mimic low-quality images with signal noise, we further add noise on the log scale simulated from a mean-zero normal distribution with standard deviation $\tau=0.1,0.5$. That is, the simulated data for each replication ($r=50$) is 
\begin{equation}
\label{eq:noise}
\log S_{mv}^{(r)}=\log \bar{S}_{mv}+E_{mv},\quad E_{mv}\sim\mathcal{N}(0,\tau^2),
\end{equation}
where $\log S_{mv}^{(r)}$ is the simulated signals for each replication ($r$), $\log S_{mv}$ is the logarithm signal from the example data, and $E_{mv}$ is simulated noise.

\subsection{Simulation Details}
We construct $N(v)$ as described before. We use the posterior mean estimate of \code{SpDiST} to compare with the estimates of alternatives. We compute posterior mean of $\bm{A}_v$ based on $2000$ MCMC samples after $3000$ samples as burn-in. In comparison, we compare our method to \code{DiST}. In addition, we also compare our method to a non-spatial method: the least squares method \citep{niethammer2006diffusion}. The least squares method \citep{niethammer2006diffusion} is to estimate $\bm{A}_v$ via $$\arg\min_{\bm{A}_v}\sum_m||\log S_{mv}-\log S_{0v}-b\bm{g}_m^T\bm{A}_v\bm{g}_m||^2.$$ For \code{DiST}, the estimates are the principal eigenvectors. To compare to \code{DiST}, for \code{SpDiST}, we compute the principal eigenvectors of the posterior means of diffusion tensor. For comparison, we also compute the principal eigenvector of the diffusion tensor estimate of the least squares method.

\subsection{Results}
To quantify the performance of the three methods, we introduce two metrics. For voxels with fiber directions, we use \textit{Metric 1} $$d_1(\bm{m}_v,\hat{\bm{m}}_v)=\text{arccos}(|\bm{m}_v^T\hat{\bm{m}}_v|),$$ a metric measuring acute angle between true tensor direction $\bm{m}_v$ and estimated $\hat{\bm{m}}_v$. A small $d_1(\bm{m}_v,\hat{\bm{m}}_v)$ indicates that the fiber direction is estimated accurately. We also introduce \textit{Metric 2} measuring the difference between true between-neighbor angle and estimated between-neighbor angle: $$d_2(\hat{\bm{m}}_v,\hat{\bm{m}}_u)=|\text{arccos}(|\hat{\bm{m}}_v^T\hat{\bm{m}}_u|)-\text{arccos}(|\bm{m}_v^T\bm{m}_u|)|,$$ where $u,v$ are neighbors. A small $d_2(\bm{m}_v,\hat{\bm{m}}_u)$ leads to an accurate decision if two voxels belong to the same fiber.

We summarize the results in Table \ref{tab:result}, including the mean estimates by averaging over $50$ replications and the associated standard errors (in parentheses). From the result, we find that \code{SpDiST} and \code{DiST} have an overall better performance in comparison to the non-spatial method. From Table \ref{tab:result}, the \code{SpDiST} is more robust to noise, which may motivate a study on the robustness of tensor direction estimates based on different parameterization. However, the noise may have little effect on \textit{Metric 2}, leading to the same fiber tracking results. Although the \code{SpDiST} and \code{DiST} have similar performance, however, the MCMC-based \code{SpDiST} provides a means to quantify the uncertainties of fiber tracking, unlike \code{DiST}.

\begin{table}[ht!]
\caption{Summary of simulation results based on \textit{Metric 1} and \textit{Metric 2}. The mean estimates by averaging over $50$ replications and the associated standard errors (in parentheses) are summarized. \label{tab:result}}
\centering
\begin{tabular}{ccccc}
\hline
Metric & Noise ($\tau$) & \begin{tabular}[c]{@{}c@{}}Least \\ Square\end{tabular} & \code{SpDiST} & \code{DiST} \\
\hline
\multirow{2}{*}{$d_1$}&  \textit{0.1} & 0.09(0.007) & 0.08 (0.006) &  0.08(0.010) \\
 & \textit{0.5} & 0.20(0.04) & 0.12(0.010)  & 0.19(0.020)\\ \hline
 \multirow{2}{*}{$d_2$} & \textit{0.1} & 0.06 (0.005) & 0.06(0.014) & 0.06(0.012) \\
 & \textit{0.5} & 0.24(0.05) & 0.08(0.010)  & 0.09(0.015)\\
 \hline
\end{tabular}
\end{table}

\section{Discussion}
\label{sec:diss}
In the numerical study, we find that \code{DiST} and \code{SpDiST} have similar performances. However, the MCMC-based \code{SpDiST} provides a probabilistic means to quantify the result of fiber tracking. This provides some potentially important information for neuroscientists to understand brain anatomical connection. Furthermore, we also give a sensitivity analysis to the tuning parameter $C$, addressing the issue raised by Kang and Li \citep{kang2016discussion}.

Although the current methodologies might be sufficient for preliminary fiber tracking, there are still several issues. One problem is that the current methods focus on developing an imaging processing tool but not on scientifically and statistically explaining the outcomes \citep{lazar2016discussion}. However, proposing a statistical approach which characterizes factors affecting the outcomes might be critical in further studies, providing more insightful information in neuroscience. However, this is challenging because to incorporate covariates in the model and to properly combine the model to a fiber tracking algorithm are not straightforward. Another issue is crossing fibers. That is the single tensor model \citep{mori1999three} fails to account for voxels where there are multiple fibers. Although it is assumed that increasing the resolution of the image may handle this issue, Schilling et al. \citep{schilling2017can} give an unexpected result that increasing the resolution is not a solution. This needs to be rigorously studied with close interdisciplinary collaboration.

\end{document}